%% file: main.tex
\documentclass[journal,twoside,web]{ieeecolor}
\usepackage{tmi}
\usepackage{cite}
\usepackage{amsmath,amssymb,amsfonts}
\usepackage{algorithmic}
\usepackage{graphicx}
\usepackage{textcomp}
\usepackage{color}
\usepackage{xspace}
\usepackage{graphicx}
\usepackage{caption}
\usepackage{booktabs}
\usepackage{multirow}
\usepackage{hyperref}
\hypersetup{colorlinks=true,
            linkcolor=subsectioncolor,
            anchorcolor=blue,
            citecolor=blue}
\usepackage[T1]{fontenc}
\usepackage{float}
\usepackage{latexsym}
\usepackage{soul}
\usepackage{makecell}

\newcommand{\methodname}{RISE-MAR\xspace}
\newcommand{\methodnameli}{RISE$^\dagger$-MAR\xspace}
\newcommand{\methodnamemali}{RISE$^{\dagger\dagger}$-MAR\xspace}
\newcommand{\papertitle}{Radiologist-in-the-Loop Self-Training for Generalizable CT Metal Artifact Reduction}

\newcommand{\bestres}[1]{\textbf{#1}}
\newcommand{\secondres}[1]{\underline{#1}}
\newcommand{\eg}{e.g.\xspace}

\newcommand{\ie}{i.e.\xspace}

\newcommand{\etc}{etc.\xspace}

\newcommand{\etal}{\emph{et al.}\xspace}
\newcommand{\vct}[1]{\boldsymbol{#1}} 
\newcommand{\mat}[1]{\boldsymbol{#1}} 

\newlength\savewidth



\def\BibTeX{{\rm B\kern-.05em{\sc i\kern-.025em b}\kern-.08em
    T\kern-.1667em\lower.7ex\hbox{E}\kern-.125emX}}
\begin{document}
\bstctlcite{IEEEexample:BSTcontrol}
\title{\papertitle}

\author{
Chenglong Ma, 
Zilong Li,~\IEEEmembership{Graduate Student Member, IEEE}, Yuanlin Li, Jing Han, 
Junping Zhang,~\IEEEmembership{Senior Member, IEEE},
Yi Zhang,~\IEEEmembership{Senior Member, IEEE},
Jiannan Liu, 
Hongming Shan,~\IEEEmembership{Senior Member, IEEE}
\thanks{\emph{Corresponding authors: Jiannan Liu, Hongming Shan.}}
\thanks{Chenglong Ma and Hongming Shan are with Institute of Science and Technology for Brain-inspired Intelligence, MOE Frontiers Center for Brain Science, and Key Laboratory of Computational Neuroscience and Brain-Inspired Intelligence, Fudan University, Shanghai 200433, China.}
\thanks{Zilong Li and Junping Zhang are with Shanghai Key Laboratory of Intelligent Information Processing, School of Computer Science, Fudan University, Shanghai 200433, China. } 
\thanks{Yuanlin Li, Jing Han, and Jiannan Liu are with Department of Oral Maxillofacial Head and Neck Oncology, Shanghai Ninth People’ s Hospital, Shanghai Jiao Tong University School of Medicine; College of Stomatology, Shanghai Jiao Tong University; National Center for Stomatology; National Clinical Research Center for Oral Diseases, Shanghai Key Laboratory of Stomatology and Shanghai Research Institute of Stomatology, Shanghai 200011, China.}
\thanks{Yi Zhang is with School of Cyber Science and Engineering, Sichuan University, Chengdu, Sichuan 610065, China. } 
}

\maketitle

\begin{abstract}
Metal artifacts in computed tomography (CT) images can significantly degrade image quality and impede accurate diagnosis. 
Supervised metal artifact reduction (MAR) methods, trained using simulated datasets, often struggle to perform well on real clinical CT images due to a substantial domain gap. 
Although state-of-the-art semi-supervised methods use pseudo ground-truths generated by a prior network to mitigate this issue, their reliance on a fixed prior limits both the quality and quantity of these pseudo ground-truths, introducing confirmation bias and reducing clinical applicability. 
To address these limitations, we propose a novel \underline{r}adiologist-\underline{i}n-the-loop \underline{se}lf-training framework for MAR, termed \methodname, which can integrate radiologists' feedback into the semi-supervised learning process, progressively improving the quality and quantity of pseudo ground-truths for enhanced generalization on real clinical CT images.
For quality assurance, we introduce a clinical quality assessor model that emulates radiologist evaluations, effectively selecting high-quality pseudo ground-truths for semi-supervised training. 
For quantity assurance, our self-training framework iteratively generates additional high-quality pseudo ground-truths, expanding the clinical dataset and further improving model generalization.
Extensive experimental results on multiple clinical datasets demonstrate the superior generalization performance of our \methodname over state-of-the-art methods, advancing the development of MAR models for practical application. Code is available at \url{https://github.com/Masaaki-75/rise-mar} . 
\end{abstract}


\input{secs/1_intro}

\input{secs/2_method}

\input{secs/3_exp}

\input{secs/4_concl}


\end{document}

%% file: secs/1_intro.tex
\section{Introduction}\label{sec:intro}

\IEEEPARstart{C}{omputed} tomography (CT) is a cornerstone in medical imaging, providing a non-invasive and high-resolution means of visualizing the detailed inner structure of the human body, at the cost of exposure to ionizing radiation that needs careful dose management~\cite{mccollough2006ct}. However, metallic implants, such as orthopedic screws, plates, dental fillings, and pacemakers, to name just a few, lead to photon starvation, beam hardening, and scattering effects~\cite{gjesteby2016metal}, 
resulting in pronounced artifacts in CT images. These artifacts manifest as bright or dark streaking and bands surrounding the implant and exhibit extreme intensities, significantly obscuring important anatomical structures and compromising diagnostic accuracy~\cite{gjesteby2016metal,kilby2002tolerance}. Therefore, reducing metal artifacts in clinical CT images remains a crucial yet challenging task.  

Conventional methods for metal artifact reduction (MAR) typically identify metal traces in the sinogram and employ interpolation to rectify the artifact-affected data~\cite{kalender1987li,meyer2010nmar}. However, they often struggle with severe artifacts that create large and complex metal traces in the sinogram, which are difficult to correct and often lead to strong secondary artifacts in the CT image. In recent years, deep learning-based approaches have revolutionized the field of MAR, demonstrating enhanced performance by leveraging complex data-driven priors from large-scale training datasets~\cite{wang2020deep}. 
Existing learning-based MAR methods can be roughly divided into supervised, unsupervised, and semi-supervised categories based on their reliance in paired data; \ie, artifact-affected images matched with corresponding artifact-free ground-truths at a pixel-level alignment. 

\begin{figure*}[t]
    \centering
    \includegraphics[width=.95\linewidth]{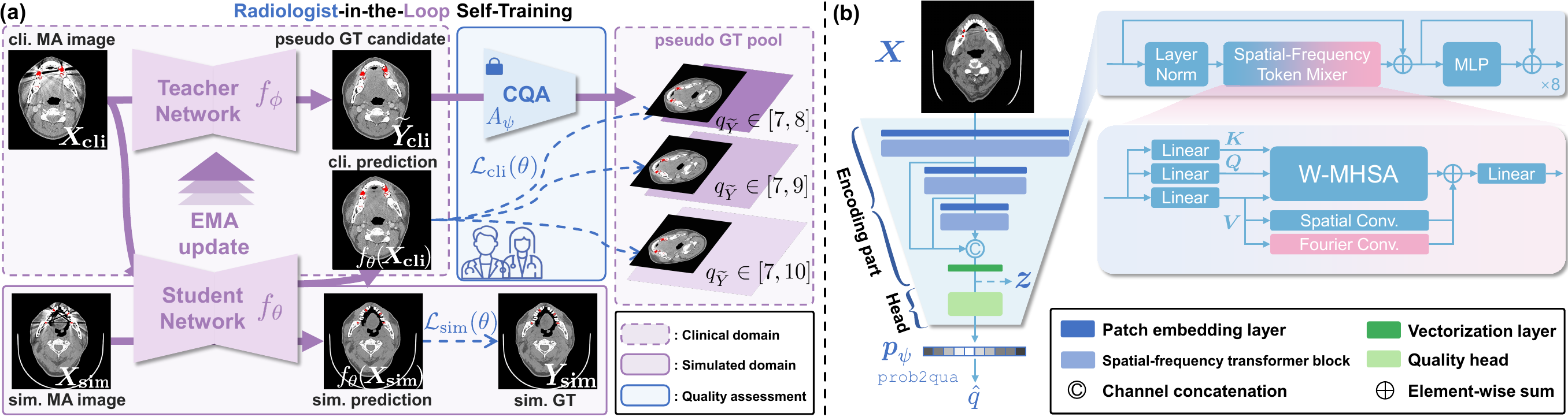}
    \caption{Overview of \methodname. (a) Radiologist-in-the-loop framework with a pretrained clinical quality assessor (CQA). (b) Overall architecture of CQA. ``MA'': metal artifact-affected, ``GT'': ground-truth. ``sim.'': simulated, ``cli.'': clinical.}
    \label{fig:method}
    \vspace{-6pt}
\end{figure*}
Supervised MAR methods~\cite{lin2019dudonet,choi2024dual,li2024quadnet} require precisely-paired data for learning to restore local structures in the CT image. 
However, such demand is impractical in real-world scenarios, prompting researchers to use metal artifact simulation techniques to create paired data for developing supervised MAR methods. Unfortunately, disparities between simulated and clinical scan settings (\eg, X-ray energy, scan geometry, \etc)~\cite{liao2019adn,xie2023dense} result in a distribution shift in metal artifacts across both the sinogram and image domains~\cite{du2021investigation}. As a result, supervised models trained with such simulated data struggle to perform well on real clinical data. 

Unsupervised MAR methods~\cite{liao2019adn,xie2023dense,wang2021bcyclegan,wang2023building,liu2024dudodp} use unpaired clinical data for training to narrow this domain gap, and demonstrate better generalization to CT images affected by real-world metal artifacts~\cite{liao2019adn}, compared to supervised methods.
Previous works focus on disentangling metal artifacts and anatomical contents in the latent space~\cite{liao2019adn,xie2023dense} or CT image space~\cite{wang2021bcyclegan,wang2023building} using generative adversarial networks (GANs) and enforcing cycle consistency. 
While such methods reduce visible artifacts, they often alter other anatomical regions in the image~\cite{du2021investigation}. This is attributed to the coarse-grained, semantic-level supervision used in unsupervised methods, as opposed to the fine-grained, pixel-level supervision employed in supervised methods. Consequently, unsupervised disentanglement methods may compromise the integrity of clinical details~\cite{wang2022idol}, particularly when complicated artifacts present~\cite{du2021investigation}. Moreover, the incorporation of multiple GAN-based loss functions introduces notorious training instability~\cite{wgan,yasin2018gantrain}.

Semi-supervised MAR methods~\cite{lyu2021ududonet,du2023udamar,wang2023semimar} have achieved a balance between preservation of structural details and generalization to clinical data, by combining the strengths of both supervised and unsupervised approaches. For example, 
UDAMAR~\cite{du2023udamar} utilizes the binary domain labels to distinguish between simulated and clinical artifacts and trains a MAR network using domain-invariant feature learning technique~\cite{dann}. Nonetheless, it discards valuable pixel-level information from unpaired clinical images, limiting the effectiveness and leaving substantial room for improvement. 
U-DuDoNet~\cite{lyu2021ududonet} and SemiMAR~\cite{wang2023semimar} integrate prior MAR networks pretrained on simulated data into the framework to produce pseudo ground-truths, offering more informative guidance on clinical detail preservation. However, the prior network remains \emph{fixed} throughout the training, which presents following limitations. First, the reliability of pseudo ground-truths generated by the prior network remains uncertain. Directly using these unverified pseudo ground-truths to train MAR models can induce confirmation bias~\cite{arazo2020confirm}. Second, the lack of adaptability in the fixed prior network results in limited diversity and quantity of high-quality pseudo ground-truths, restricting the model’s capacity to fully generalize to the clinical domain. 

To further mitigate the domain gap and limitations of current semi-supervised methods, it is critical to ensure both the \emph{quality} and \emph{quantity} of pseudo ground-truths for model training and generalization. We decompose this problem into two iterative steps: 
(1) \emph{Quality assurance}, which leverages radiologists' feedback to select those high-quality MAR predictions as pseudo ground-truths for further training, preventing confirmation bias from low-quality MAR results; and (2) \emph{Quantity assurance}, which progressively increases the number of high-quality pseudo ground-truths for effective MAR training by dynamically updating the (prior) MAR model's knowledge on the clinical domain.

To this end, we introduce \methodname, a radiologist-in-the-loop self-training framework for generalizable clinical MAR.
Specifically, \methodname comprises two key components: the clinical quality assessor (CQA) model for quality assurance and the self-training for quantity assurance. 
First, we train a CQA to provide radiologist-aligned feedback on whether a CT image qualifies for clinical use. This helps filtering out low-quality artifact-reduced images that could propagate errors if used as pseudo ground-truths for semi-supervised training. The pretrained CQA can also serve as an evaluation tool for clinical MAR task. 
Second, we propose a self-training framework that can leverage the high-quality pseudo ground-truths---assured by CQA---to update the clinical-domain knowledge of a student MAR network and a teacher MAR network, thereby bootstrapping more pseudo ground-truths and expanding the pool of clinical domain supervision. 
Together, \methodname maximizes the utility of paired simulated CT images and unpaired clinical CT images, progressively transferring the knowledge from simulated data to clinical data and enhancing the model’s generalization to clinical MAR tasks.

Our contributions can be summarized as follows.
\begin{itemize}
    \item We present \methodname, a novel radiologist-in-the-loop self-training framework for clinical MAR generalization. To the best of our knowledge, this is the first work to incorporate the radiologists' feedback for the clinical MAR task. 
    \item We propose to construct a clinical quality assessor that provides efficient and radiologist-aligned quality assessments to select high-quality pseudo ground-truths, improving the reliability of clinical-domain supervision. 
    \item We propose a novel self-training framework that dynamically bootstraps the generation of pseudo ground-truths, expanding the pool of clinical domain supervision. 
    \item Extensive experimental results on multiple datasets covering different anatomies show the superior performance of \methodname on clinical data over state-of-the-art methods. 
\end{itemize}

The rest of this paper is structured as follows. In Sec.~\ref{sec:method}, we elaborate on the proposed \methodname, followed by detailed experimental setup and results in Sec.~\ref{sec:exp}. Finally, we conclude with a discussion and final remarks in Sec.~\ref{sec:concl}.

%% file: secs/2_method.tex
\section{Methodology}\label{sec:method}
Fig.~\ref{fig:method}(a) presents the proposed framework, \methodname, designed for generalizable CT metal artifact reduction in clinical scenarios. \methodname facilitates a dynamic interaction between a radiologist-aligned clinical quality assessor model (Sec.~\ref{sec:method-cqa}) and a self-training framework (Sec.~\ref{sec:method-selftrain}). 
The clinical quality assessor (CQA) ensures quality assurance by filtering out low-quality pseudo ground-truth candidates, while the self-training framework leverages the high-quality pseudo ground-truths to iteratively train and refine MAR networks. Through this interaction, \methodname progressively transfers knowledge from simulated data to clinical data.

Concretely, given a real metal artifact-affected CT image $\mat{X}_\mathrm{cli}$, a teacher MAR network $f_\phi$ pretrained on simulated datasets predicts an initial MAR result $\widetilde{\mat{Y}}_\mathrm{cli}=f_\phi(\mat{X}_\mathrm{cli})$, which is then evaluated by the pretrained CQA to determine its quality as pseudo ground-truth for subsequent training phases. Further, the self-training framework leverages the verified pseudo ground-truths to construct ``paired'' clinical data and combines them with simulated data $(\mat{X}_\mathrm{sim}, \mat{Y}_\mathrm{sim})$ to train the student MAR network $f_\theta$ and update the teacher MAR network $f_\phi$, effectively bootstrapping more high-quality pseudo ground-truths. 

In the following, we detail these two key designs.

\subsection{Clinical Quality Assessor}
\label{sec:method-cqa}
To alleviate the confirmation bias when using pseudo ground-truths in semi-supervised learning, one might involve some radiologists in the MAR network training to assess the clinical quality of pseudo ground-truths continuously. 
Such approach, however, is impractical due to the significant time and effort required. In this paper, we offer a more feasible solution: distilling the feedback from radiologists into a CQA model to enable efficient and radiologist-aligned evaluation in the semi-supervised learning loop.

\subsubsection{Clinical Quality Assessment Collection}
\label{sec:method-cqa-dataset}
To establish a CQA that aligns with radiologists' expertise, we constructed a clinical quality assessment dataset that comprises CT images and associated quality scores annotated by two experienced radiologists. 
We start by collecting diverse clinical CT images covering torso and dental regions; these images encompass a variety of sources, anatomies, and levels of metal artifact corruption. Please refer to Sec.~\ref{sec:exp-dataset-cqa} for more details. 

Next, two experienced radiologists evaluate the \emph{clinical quality} of these images based on the severity of metal artifacts and tissue distortion. The ``clinical quality'' refers to the degree to which a CT image offers sufficient clarity and anatomical integrity for reliable interpretation by radiologists. Each image is rated on a ten-point Likert scale, where a quality of 1 signifies severe distortion in tissues within the region of interest, and a quality of 10 indicates an artifact-free image with well-preserved anatomical integrity and sharpness that is suitable for clinical diagnosis.

Specifically, for paired CT images with artifact-free ground-truth, we provide the radiologists with absolute pixel errors in the region of interest, excluding the metal area. By analyzing multiple image pairs along with their error information, the radiologists establish a ten-class classification rule (\eg, a piecewise function) which is later applied to rate image quality based on deviations from the ground-truth. This approach allows for efficient quality annotation of paired CT images. 
For unpaired CT images, the radiologists annotate the quality based on visual assessment and their clinical experience.
For each image, one radiologist assigns an initial quality score, which is subsequently reviewed by another senior radiologist. Eventually, they reach a consensus to give the final annotation.

\subsubsection{Diverse Quality Augmentation}
\label{sec:method-cqa-dqaug}
The clinical assessment dataset allows us to train a CQA that reflects radiologists' preference on CT images. Note that the CQA should not only distinguish between low- and high-quality CT images but also identify subtle variations within moderate-quality, artifact-reduced CT images processed by MAR algorithms. These images, albeit with less metal artifacts compared to low-quality ones, may still exhibit novel artifacts and distortions. 
To this end, we propose Diverse Quality Augmentation (DQAug), a strategy to enhance the CQA model by creating a diverse range of image qualities within the paired data in our clinical quality assessment dataset.

DQAug involves two steps. First, under-trained MAR algorithms are applied to low-quality artifact-affected CT images in the clinical quality assessment dataset, producing moderate-quality images. Then, DQAug employs MixUp~\cite{zhang2018mixup} data augmentation across low, moderate, and high-quality images, blending them to create new images that significantly increase the diversity of the dataset. These strategies are employed in an on-the-fly manner during the CQA training process, ensuring that the CQA is continuously exposed to a wide spectrum of image qualities and leading to a more robust model.

\subsubsection{Aligning CQA to Radiologists' Feedback}
Since the quality assessment data annotated by the radiologists are on a ten-point Likert scale, we formulate the alignment as a ten-class classification problem. As shown in Fig.~\ref{fig:method}(b), we build CQA by adapting the encoder of a generic medical image reconstruction network~\cite{ma2023proct}, which demonstrates strong capabilities of feature extraction for CT images in our experiments. 

As shown in Fig.~\ref{fig:method}(b), CQA is a transformer-based multi-scale encoder with a trailing vectorization layer, and a quality head consisting of a multi-layer perceptron (MLP) followed by a softmax layer. Each scale begins with a patch embedding layer that converts the image into overlapping patch embedding, followed by eight spatial-frequency transformer blocks. The spatial-frequency token mixer within each transformer block consists of a windowed multi-head self-attention (W-MHSA) layer~\cite{liu2021swin} and two convolution layers---one operating in the spatial domain and the other in the frequency domain. The feature maps from three scales are resized and concatenated to form the multi-scale feature maps. 

\begin{figure}[t]
    \centering
    \includegraphics[width=\linewidth]{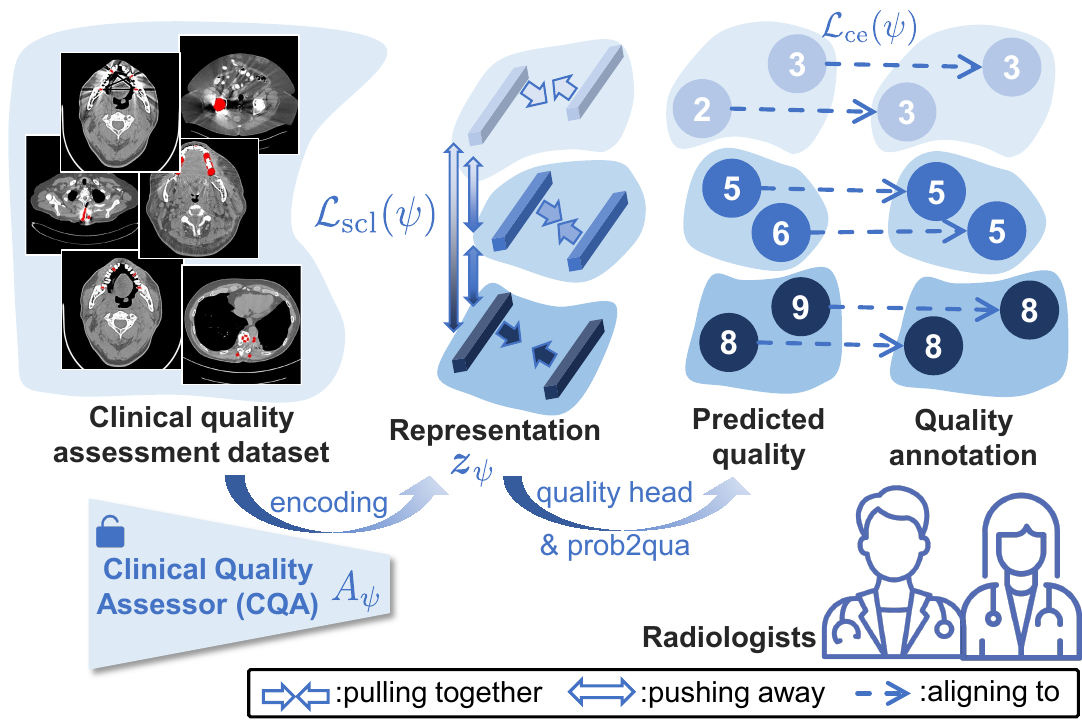}
    \caption{Training of CQA. The predicted probability vector $\vct{q}_\psi$ is omitted for more intuitive presentation.}
    \label{fig:cqa-training}
    \vspace{-12pt}
\end{figure}

Let $A_\psi$ denote our CQA model parameterized by $\psi$. Given a CT image $\mat{X}$, CQA extracts the multi-scale feature maps, which will be passed through a vectorization layer that performs pooling, flattening, and normalization, thereby producing latent vector $\vct{z}_\psi$. The latent vector is processed by the quality head to obtain the predicted probability vector $\vct{p}_\psi=A_\psi(\mat{X})\in\mathbb{R}^{10}$. The predicted quality $\widehat{q}$ is then given by the following \texttt{prob2qua} function:
\begin{equation}
    \widehat{q} = \texttt{prob2qua}(\vct{p}_\psi) := 
    \sum_{k=1}^{10} k\vct{p}_\psi^{(k)}, \quad \widehat{q}\in [1,10],
\label{eq:prob2qua}
\end{equation}
where $\vct{p}_\psi^{(k)}$ denotes the $k$-th element of $\vct{p}_\psi$. We train CQA with a compound loss function $\mathcal{L}_\mathrm{cqa}$, which consists of a cross-entropy loss $\mathcal{L}_\mathrm{ce}$ and a supervised contrastive loss $\mathcal{L}_\mathrm{scl}$, as depicted in Fig.~\ref{fig:cqa-training}. Specifically, let $\vct{p}$ denote the one-hot quality label. $\mathcal{L}_\mathrm{ce}$ is defined as:
\begin{align}
    \mathcal{L}_\mathrm{ce}(\psi) &= - \vct{p} \log \vct{p}_\psi = - \vct{p} \log A_\psi(\mat{X}).
\label{eq:cqa-loss-ce}
\end{align}
Besides, we introduce $\mathcal{L}_\mathrm{scl}$ to better distinguish the subtle quality difference between CT images~\cite{li2023gloredi}. This loss encourages representations with the same quality label to cluster together while pushing apart those with different labels. Let $\vct{z}_\psi$ be the latent vector representing the current input image labeled with a quality of $q$, $M$ a fixed-length memory bank that stores latent vectors from previous image batches over the training steps, and $M_q\subset M$ the set of latent vectors that encode previous images labeled as $q$. Then, $\mathcal{L}_\mathrm{scl}$ is defined as:
\begin{align}
    \mathcal{L}_\mathrm{scl}(\psi) &= - \dfrac{1}{|M_q|}
    \sum_{\vct{z}_{j}\in M_q}\log \dfrac{
    \exp(\vct{z}_\psi^ \top \vct{z}_{j} / \tau)}
    {\sum_{\vct{z}'\in M} 
    \exp(\vct{z}_\psi^\top\vct{z}'/ \tau)},
\label{eq:cqa-loss-ctr}
\end{align}
where $|M_q|$ denotes the cardinality of $M_q$. The memory bank $M$ keeps updating in a first-in-first-out manner during training, when its capacity is reached ($|M|=300$). The compound loss $\mathcal{L}_\mathrm{cqa}(\psi)$ for training CQA model $A_\psi$ is given by:
\begin{align}
    \mathcal{L}_\mathrm{cqa}(\psi) &= \mathcal{L}_\mathrm{ce}(\psi) + \lambda_\mathrm{scl} \mathcal{L}_\mathrm{scl}(\psi),
\label{eq:cqa-loss}
\end{align}
where the factor $\lambda_\mathrm{scl}$ is empirically set to 0.01 to balance the magnitude of the two loss terms.

\begin{table*}[!htb]
\centering
\caption{Overview of datasets. ``sim.-to-cli.'': simulation-to-clinical; ``sim.-to-sim.'': simulation-to-another-simulation.}
\resizebox{\linewidth}{!}{
\begin{tabular}{lcccccc}
\toprule
Datasets &Sources & Regions & Data Format & Paired Data? & With GT? & Usage \\ 
\midrule
Simulated DeepLesion &DeepLesion~\cite{yan2018deeplesion} & torso &CT images &$\checkmark$ &$\checkmark$ & In-domain MAR validation \\
Clinical DeepLesion &DeepLesion~\cite{yan2018deeplesion} &torso &CT images &  &  & Out-of-domain MAR (sim.-to-cli.) \\
\midrule
Simulated CTPelvic1K &CTPelvic1K~\cite{liu2021ctpelvic} & torso &CT images &  &$\checkmark$ & Out-of-domain MAR (sim.-to-sim.) \\
\midrule
Simulated Dental &Private &dental &CT images &$\checkmark$ &$\checkmark$ & In-domain MAR validation \\
Clinical Dental &Private & dental &CT images &  &  & Out-of-domain MAR (sim.-to-cli.) \\
\midrule
\makecell[l]{Clinical Quality \\ Assessment} 
& \makecell{DeepLesion~\cite{yan2018deeplesion} \\ \& CTPelvic1K~\cite{liu2021ctpelvic} \& Private} 
& \makecell{torso \\ \& dental} 
& \makecell{CT image with \\ quality score} 
& $\checkmark$ & $\checkmark$ & Development of CQA model \\
\bottomrule
\end{tabular}}
\vspace{-8pt}
\label{tab:dataset-overview}
\end{table*}

\subsection{Radiologist-in-the-Loop Self-Training}
\label{sec:method-selftrain}
The radiologist-aligned CQA can serve as a quality filter in the training of MAR networks, ensuring that only ``pseudo ground-truth candidates'' meeting predefined quality requirement are utilized. However, relying on a fixed prior model~\cite{lyu2021ududonet,wang2023semimar} for generating such candidates may hinder generalization, as they may not be consistently accepted by CQA, thus limiting the pool of high-quality pseudo ground-truths. 
To address this, we propose a self-training framework that involves the dynamic interaction between a teacher MAR network and a student MAR network, progressively increasing the number of viable pseudo ground-truths and enhancing MAR in clinical CT images, as shown in Fig.~\ref{fig:method}(a). 

Both the teacher network $f_\phi$ and the student network $f_\theta$ share the same architecture, but the teacher $f_\phi$ is equipped with prior MAR knowledge by pretraining on simulated data, while the student $f_\theta$ is initialized as an under-trained network. Through the CQA quality assurance, $f_\phi$ can provide high-quality pseudo ground-truths for $f_\theta$ to learn the MAR in the clinical domain. In turn, $f_\theta$ updates the clinical-domain knowledge of $f_\phi$ using exponential moving average (EMA), enabling an increasing number of predictions from $f_\phi$ to pass the CQA and ensuring a robust supply of data for clinical-domain supervision. The interaction between $f_\phi$ and $f_\theta$ is a combination of two steps: a simulated-domain supervised learning step and a clinical-domain unsupervised learning step.

\subsubsection{Simulated-Domain Supervised Learning}
This step leverages paired CT images, \ie, a simulated artifact-affected image $\mat{X}_\mathrm{sim}$ and its corresponding artifact-free ground-truth $\mat{Y}_\mathrm{sim}$, to train the student MAR network $f_\theta$, learning to restore the local structures in the image with a pixel-wise $\ell_1$ loss: 
\begin{align}
    \mathcal{L}_\mathrm{sim}(\theta) &= \Vert f_\theta(\mat{X}_\mathrm{sim}) - \mat{Y}_\mathrm{sim}
    \Vert_1.
\label{eq:loss-mar-sim}
\end{align}

\subsubsection{Clinical-Domain Unsupervised Learning}
This step exploits unpaired clinical CT images to train the student MAR network $f_\theta$, enhancing its generalizability to real clinical data. From each unlabeled, artifact-affected clinical image $\mat{X}_\mathrm{cli}$, the teacher network $f_\phi$ first predicts a pseudo ground-truth $\widetilde{\mat{Y}}_\mathrm{cli} = f_\phi(\mat{X}_\mathrm{cli})$. This pseudo ground-truth is then evaluated by our radiologist-aligned CQA model $A_\psi$ to determine its quality $q_{\widetilde{Y}}$ as follows: 
\begin{align}
    q_{\widetilde{Y}} = \texttt{prob2qua}(A_\psi(\widetilde{\mat{Y}}_\mathrm{cli})).
\end{align}
When $q_{\widetilde{Y}}$ falls outside a predefined quality range $\mathcal{Q}$, it indicates that $\widetilde{\mat{Y}}_\mathrm{cli}$ does not qualify as a pseudo ground-truth due to potential inaccuracies. Consequently, we exclude such cases from the training to prevent confirmation bias. 
On the contrary, if $q_{\widetilde{Y}}$ lies within $\mathcal{Q}$, then $\widetilde{\mat{Y}}_\mathrm{cli}$ is deemed valuable for guiding $f_\theta$ in removing real metal artifacts in CT images. 

With the verified pseudo ground-truth, we treat the residue $\mat{R}_\mathrm{cli} = \mat{X}_\mathrm{cli} - \widetilde{\mat{Y}}_\mathrm{cli}$ as ``artifact'' in $\mat{X}_\mathrm{cli}$, thereby creating a set of paired data by combining another unpaired artifact-free CT image $\mat{Y}'_\mathrm{cli}$, denoted by $\mathcal{C}=\{ (\mat{X}_\mathrm{cli}, \widetilde{\mat{Y}}_\mathrm{cli}), (\mat{X}'_\mathrm{cli}, \mat{Y}'_\mathrm{cli}) \}$, where $\mat{X}'_\mathrm{cli}=\mat{Y}'_\mathrm{cli}+\mat{R}_\mathrm{cli}$. The pseudo pairs in $\mathcal{C}$ can then be used to train $f_\theta$ through the clinical-domain unsupervised loss defined as follows: 
\begin{align}
    \mathcal{L}_\mathrm{cli}(\theta) &= 
    \mathbb{I}_{q_{\widetilde{Y}}\in\mathcal{Q}} \cdot
    \mathbb{E}_{(\mat{X},\mat{Y})\sim\mathcal{C}}
    \Vert f_\theta(\mat{X}) - \mat{Y} \Vert_1,
\label{eq:loss-mar-cli}
\end{align}
where $\mathbb{I}_{\{\cdot\}}$ is an indicator function, and $\mathcal{Q}=[7,10]$ in this paper.

\subsubsection{Self-Training Objective Function}
Finally, the student MAR network $f_\theta$ learns to integrate the simulated-domain and clinical-domain knowledge by minimizing the following loss: 
\begin{align}
    \mathcal{L}(\theta) &= \mathcal{L}_\mathrm{sim}(\theta) + \mathcal{L}_\mathrm{cli}(\theta),  
\label{eq:loss-mar}
\end{align}
where $f_\theta$ receives supervision from both simulated and clinical data. On the other hand, $f_\phi$ updates its parameters through EMA with a decay rate $\eta=0.999$, defined as: 
\begin{align}
    \phi \leftarrow \eta \phi + (1 - \eta) \theta, 
\label{eq:ema}
\end{align}
thereby stably updating the clinical-domain MAR knowledge and bootstrapping higher-quality pseudo ground-truths.

Note that the MAR networks ($f_\theta$ and $f_\phi$) within our self-training framework can be any effective learning-base MAR models. The self-training guided by radiologist-aligned quality assessment from our CQA model constructs the proposed radiologist-in-the-loop self-training framework, \methodname.

%% file: secs/3_exp.tex
\section{Experiments}\label{sec:exp}

\subsection{Datasets}\label{sec:exp-dataset}
We use the following CT image dataset sources for MAR model training and evaluation: (1) DeepLesion dataset~\cite{yan2018deeplesion}, a large multi-source publicly available dataset with clinical images of the central body area, (2) CTPelvic1K dataset~\cite{liu2021ctpelvic}, another public clinical dataset focusing on the pelvic regions, and (3) Dental dataset, a private dataset collected from maxillofacial CT scanners at Shanghai Ninth People’s Hospital. 
For each of them, simulated data are created from the artifact-free clinical CT images. 
Additionally, we create a clinical quality assessment dataset to train and test the CQA model. All images in the experiments have a resolution of 512 $\times$ 512. Please refer to Table~\ref{tab:dataset-overview} for an overview of datasets.

\subsubsection{DeepLesion Dataset}\label{sec:exp-dataset-deepl}
We first identify artifact-free clinical CT images from DeepLesion dataset by selecting those with over 100 pixels exceeding 3,000 Hounsfield Units (HU), following UDAMAR~\cite{du2023udamar}. Then, we apply the metal artifact simulation algorithm in Zhang \etal's work~\cite{cnnmar} to simulate 7,200 training pairs using 90 predefined metal masks, and 800 test pairs using another 10 metal masks from 4,000 artifact-free images. Additionally, 1,749 artifact-free and artifact-affected images from 344 cases constitute the clinical data.

\subsubsection{CTPelvic1K Dataset}\label{sec:exp-dataset-pelvic}
For the CTPelvic1K dataset, we simulate 1,662 paired artifact-affected and artifact-free CT images from 103 patients, where slices from 92 patients are used for training and the rest 11 patients are for testing. For a given CT image, the metal mask for artifact simulation is randomly chose from either the bone segmentation mask or one of ten masks used in DeepLesion test data. The bone segmentation mask is obtained via threshold-based image segmentation with HU threshold of 900. 
Note that the simulation settings (\eg, metal shapes and densities) and the imaging manufacturers for CTPelvic1K differ from those in the Dental and DeepLesion datasets, enabling this dataset to be used as a quantitative performance tracker that allows for generalization comparison across different MAR methods.

\subsubsection{Dental Dataset}\label{sec:exp-dataset-dental}
The Dental dataset comprises 1,125 metal artifact-affected and 2,233 unpaired artifact-free maxillofacial CT images from 104 de-identified cases. We generated 7,200 training pairs and 800 test pairs from 45 cases to form the simulated data, where the metal masks are randomly extracted from the teeth segmentation results. The remaining 1,423 artifact-free images and 1,125 artifact-affected images from 59 cases constitute the clinical data.

\subsubsection{Clinical Quality Assessment Dataset}\label{sec:exp-dataset-cqa}
This dataset comprises 25,000 simulated pairs of artifact-affected and artifact-reduced CT images, along with 3,895 unpaired CT images, sourced from the above three datasets. The metals in the simulated images are of various types, shapes, and locations. As described in Sec.~\ref{sec:method-cqa-dataset}, each image has been assigned a clinical quality score by the radiologists. We split the training and test sets by the ratio of 9:1, and ensure no overlap with those test data used to evaluate the MAR methods.

\begin{table*}[!htb]
\centering
\caption{Quantitative evaluation [PSNR (db), SSIM (\%), CQA quality] of different methods on DeepLesion and CTPelvic1K test sets. The best results are highlighted in \bestres{bold} and the second-best results are \secondres{underlined}.
}
\resizebox{1.0\textwidth}{!}{
\begin{tabular}{ccccccccc} 
\toprule
{\textbf{Test sets}} & {\textbf{Metrics}} & Input 
& LI~\cite{kalender1987li} 
& Supervised~\cite{wang2021bcyclegan} 
& $\beta$-CycleGAN~\cite{wang2021bcyclegan} 
& UDAMAR~\cite{du2023udamar} 
& SemiMAR~\cite{wang2023semimar} 
& \methodname (ours) \\
\midrule
\multirow{3}[0]{*}{\shortstack[c]{Simulated DeepLesion\\(in-domain)}}  
& PSNR 
&34.51 &40.58 &\bestres{47.68} &40.50 
&45.56 &44.54 &\secondres{47.25} \\
& SSIM 
&90.49 &95.79 &\bestres{99.35} &97.56 
&99.11 &98.70 &\secondres{99.30} \\
& CQA quality
&4.846 &6.002 &\secondres{7.793} &6.022 
&7.211 &7.494 &\bestres{8.169} \\
\midrule
\multirow{3}[0]{*}{\shortstack[c]{Simulated CTPelvic1K\\(out-of-domain)}}  
& PSNR 
&38.87 &43.84 &41.67 &42.81 
&46.04 &\secondres{46.42} &\bestres{47.88} \\
& SSIM 
&98.74 &98.19 &93.57 &98.49 
&98.97 &\secondres{99.10} &\bestres{99.25} \\
& CQA quality 
&7.371 &6.849 &8.154 &6.973 
&\secondres{8.396} &8.110 &\bestres{8.645} \\
\midrule
\parbox[c][\totalheight][c]{0.15\linewidth}{\centering Clinical DeepLesion\\(out-of-domain)}
& CQA quality 
&7.809 &--\textsuperscript{*} &6.577 &6.334 
&8.132 &\secondres{8.154} &\bestres{8.554} \\
\bottomrule
\end{tabular}
}
\caption*{\raggedright\small{(*: Data unavailable due to the absence of the corresponding real sinogram, making LI infeasible.)}}
\label{tab:torso-quant}
\end{table*}
\begin{figure*}
\centerline{\includegraphics[width=.95\linewidth]{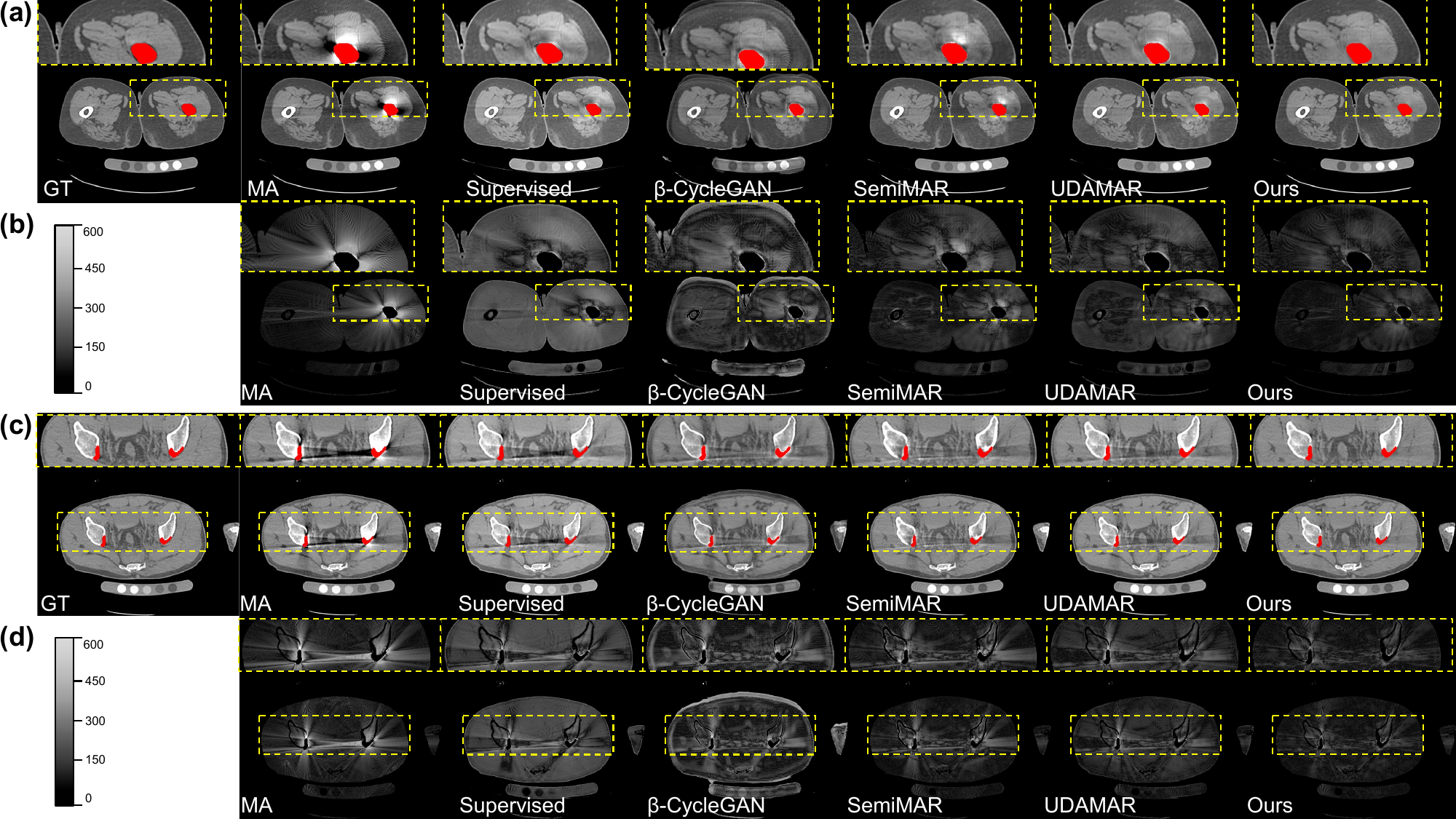}}
\caption{Visual results of different MAR methods on two typical slices from simulated CTPelvic1K test set. Sub-figures (a) and (c) are CT images, (b) and (d) are corresponding absolute error maps. For better visualization, metal pixels are filled with red color. Regions of interest are enlarged. The display window is [-300, 300] HU.}
\label{fig:pelvic12}
\end{figure*}

\subsection{Implementation Details}\label{sec:exp-impl}
We compare \methodname with a traditional sinogram-inpainting method (LI)~\cite{kalender1987li}, a supervised baseline method~\cite{wang2021bcyclegan}, and state-of-the-art unsupervised and semi-supervised MAR methods including $\beta$-CycleGAN~\cite{wang2021bcyclegan}, UDAMAR~\cite{du2023udamar}, and SemiMAR~\cite{wang2023semimar}. All learning-based models are trained using Adam optimizer with $(\beta_1, \beta_2)=(0.5, 0.999)$ on an NVIDIA 4090 GPU. 
Competing methods that utilize generative adversarial training often face training instabilities, which can significantly degrade their performance. We found that this could be mitigated by initializing the MAR network parameters with weights pretrained for 10 epochs in a supervised manner. For a fair comparison, the MAR networks involved in other methods are also initialized with the same parameters. \methodname is trained for another 30 epochs with a batch size of 1. Our CQA is also trained for 30 epochs. The initial learning rate is $10^{-4}$ and is reduced by half for every 20 epochs. Data augmentation techniques, including random flipping and rotation, are applied during training of all methods.

The MAR network used for the aforementioned learning-based methods is the attention-based U-Net proposed in $\beta$-CycleGAN~\cite{wang2021bcyclegan}, which incorporates spatial and channel attention modules into the skip connection in U-Net to enhance expressivity while maintaining computational efficiency. The discriminators in $\beta$-CycleGAN, SemiMAR, and UDAMAR are adopted from PatchGAN~\cite{zhu2017cyclegan}. 

For preprocessing, each CT image was preprocessed by clipping its HU values to the range of $[-1024, 3072]$, followed by min-max normalization to scale the values to $[0, 1]$.

\subsection{Evaluation Metrics}
\subsubsection{Metal Artifact Reduction Evaluation}
We employ peak signal-to-noise ratio (PSNR) and structural similarity (SSIM)~\cite{ssim} to evaluate the performance of different MAR methods when ground-truth data are available. We also provide MAR quality results evaluated by our CQA (CQA quality), since CQA itself can serve as an automated MAR quality assessor, which is demonstrated in Sec.~\ref{sec:exp-res-cli}. 

For clinical data without ground-truth data, the evaluation primarily involves visual assessment, with CQA quality as a supplementary quantitative metric.

\subsubsection{Assessment Accuracy Metrics}
Spearman's rank correlation coefficient (SRCC) and Pearson’s linear correlation coefficient (PLCC) are employed to evaluate the performance of our CQA; they are two metrics commonly used in ranking tasks like image quality assessment~\cite{iqagpt,liang2024rich}. Specifically, SRCC measures the correlation between the ranked values of the predictions and the corresponding ground-truths \ie, the prediction monotonicity. PLCC quantifies the linear relationship between the predicted scores and ground-truth scores, \ie, the prediction accuracy. The higher values of SRCC and PLCC (both close to 1) indicate better alignment between the prediction of CQA model and expert clinical assessments.

\begin{table*}[!t]
\centering
\caption{Quantitative evaluation [PSNR (db), SSIM (\%), CQA quality] of different methods on Dental test sets. The best results are highlighted in \bestres{bold} and the second-best results are \secondres{underlined}.
}
\resizebox{1.0\textwidth}{!}{
\begin{tabular}{lccccccccc} 
\toprule
{\textbf{Test sets}} & {\textbf{Metrics}} & Input 
& LI~\cite{kalender1987li} 
& Supervised~\cite{wang2021bcyclegan} 
& $\beta$-CycleGAN~\cite{wang2021bcyclegan} 
& UDAMAR~\cite{du2023udamar} 
& SemiMAR~\cite{wang2023semimar} 
& \methodname (ours) \\
\midrule
  & PSNR 
&31.71 &35.14 &\bestres{45.24} &40.15 
&41.32 &41.94 &\secondres{44.83} \\
Simulated Dental (in-domain) & SSIM 
&93.86 &93.60 &\bestres{99.39} &98.42 
&99.23 &98.95 &\secondres{99.34} \\
 & CQA quality
&4.607 &4.168 &7.308 &6.984 
&\secondres{7.917} &7.258 &\bestres{8.106} \\
\midrule
Clinical Dental (out-of-domain) & CQA quality 
&5.765 &--\textsuperscript{*} &6.043 &5.865 &\secondres{6.475} &6.245 &\bestres{7.304} \\
\bottomrule
\end{tabular}
}
\caption*{\raggedright\small{(*: Data unavailable due to the absence of the corresponding real sinogram, making LI infeasible.)}}
\label{tab:dental-quant}
\vspace{-12pt}
\end{table*}

\subsection{Generalization Comparison on Simulated Datasets}\label{sec:exp-res-syn}
We first compare the generalizability of \methodname and other MAR methods on the simulated datasets quantitatively. 
The training set in this experiment includes simulated data from DeepLesion and CTPelvic1K datasets. Specifically, DeepLesion dataset provides paired images, whereas the training data from CTPelvic1K dataset are randomly shuffled to be \emph{unpaired}, mimicking the scenario of training with unpaired clinical data. Unsupervised or semi-supervised MAR methods utilize training data from two datasets, while supervised baseline is trained only with paired simulated data from DeepLesion dataset. Test data from DeepLesion and CTPelvic1K datasets are used for in-domain and out-of-domain quantitative evaluation, respectively.

Despite both datasets being simulated, they significantly differ in their imaging manufacturers and artifact simulation parameters, making this experiment suitable to test the domain generalizability of MAR models quantitatively.

The results are shown in the first two groups of Table~\ref{tab:torso-quant}, namely, the groups of ``Simulated DeepLesion'' and ``Simulated CTPelvic1K''.
The results in the first group of Table~\ref{tab:torso-quant} show that when the metal artifacts in the test set have a similar distribution to those in the training set (\ie, in-domain scenario), most learning-based MAR methods significantly outperform the traditional LI interpolation and substantially enhance image quality compared to the input CT images.
Among them, supervised method achieves the best PSNR and SSIM results. 
However, it demonstrates a significant performance decline when directly applied to the simulated CTPelvic1K dataset, due to the domain gap between the two datasets. 
Unsupervised $\beta$-CycleGAN shows limited performance on both datasets while showing better generalizability on the out-of-domain simulated CTPelvic1K dataset. By comparison, semi-supervised methods, including UDAMAR, SemiMAR, and the proposed \methodname, effectively narrowing the performance gap between two datasets, where our \methodname achieves competitive results compared with the supervised method. In the out-of-domain CTPelvic1K dataset, \methodname achieves the best performance in terms of PSNR, SSIM and CQA quality, indicating that \methodname not only retains in-domain MAR knowledge but also shows strong generalizability on out-of-domain datasets.

\begin{figure}[t]
    \centering
    \includegraphics[width=\linewidth]{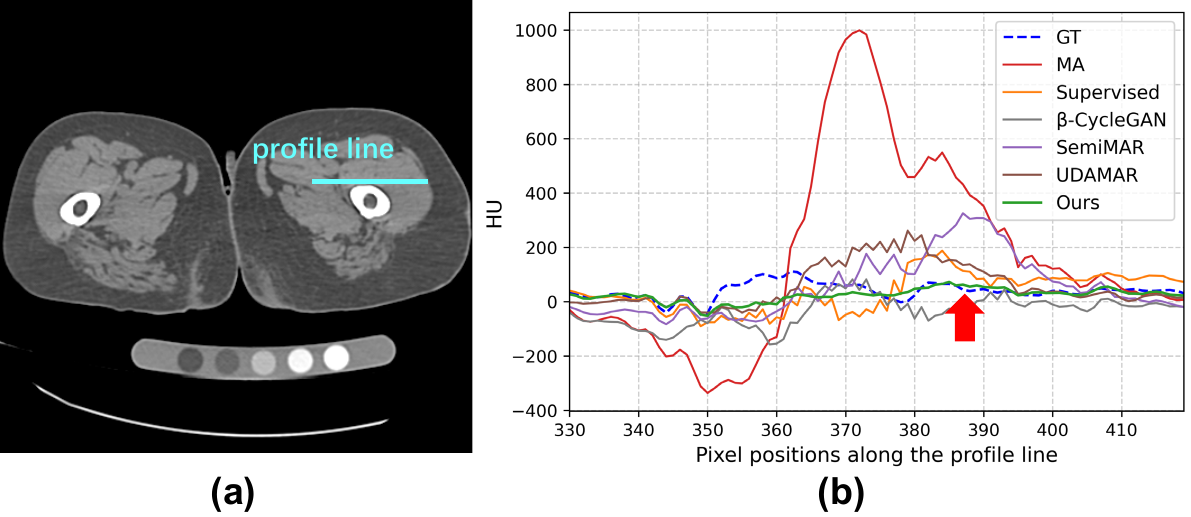}    \caption{Profile line plot comparison of MAR methods. (a) CT image with a profile line (cyan colored); (b) HU values along the profile line.}
    \label{fig:lineplot}
    \vspace{-12pt}
\end{figure}

Fig.~\ref{fig:pelvic12} shows visual comparisons on two slices from out-of-domain CTPelvic1K test set, including the predicted artifact-reduced images and the corresponding absolute error maps against artifact-free ground-truths. From the absolute error maps, we observe that the supervised baseline tends to change the overall pixel values, due to the domain gap between two simulated datasets. 
While $\beta$-CycleGAN can remove some artifacts, it also introduces spurious tissue structures, as evident in the absolute error maps, particularly in the dark red regions of the slices. This issue likely arises because $\beta$-CycleGAN struggles to fully disentangle artifacts from the underlying image content. The lack of pixel-wise alignment supervision further exacerbates this problem, leading to the loss of critical clinical details that are even less affected by metal artifacts. 

Semi-supervised methods like SemiMAR and UDAMAR provide better visual results for areas affected by the metal artifacts. However, the bright streak artifacts remain. In contrast, our proposed \methodname shows a notable improvement in restored image quality, showing minimal deviations from the ground-truth, while other methods yield larger areas of high deviation, suggesting the superiority of \methodname for effective artifact reduction.

In Fig.~\ref{fig:lineplot}, we present profile line plot comparison for the first slice in Fig.~\ref{fig:pelvic12}(a). The blue dotted curve in Fig.~\ref{fig:lineplot}(b) represents the profile of ground-truth image, while the green curve represents our \methodname The red arrow  highlights areas where our method demonstrates better alignment with the ground-truth image.

\begin{figure*}[htb]
\centerline{\includegraphics[width=.9\linewidth]{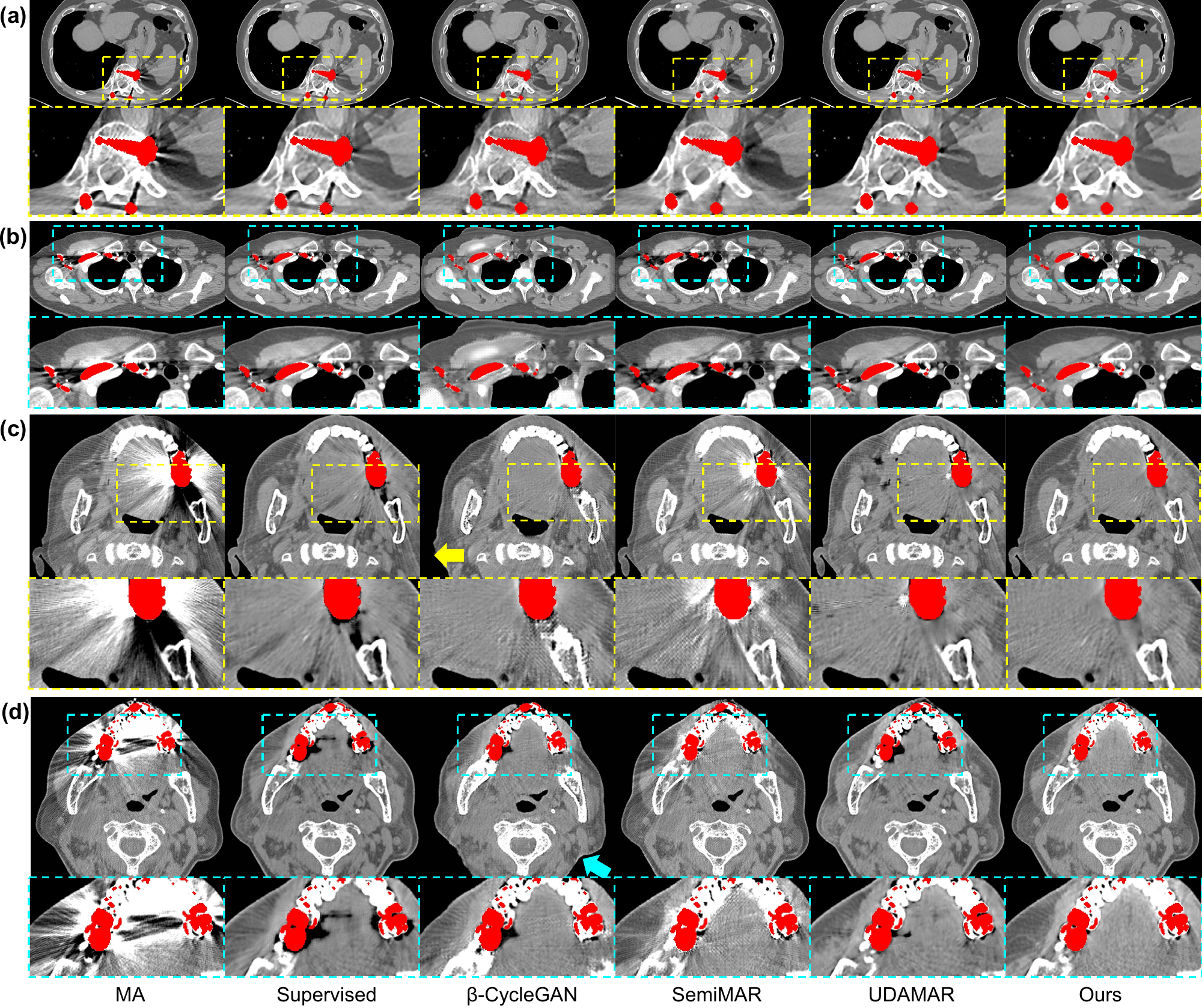}}
\caption{Visual results of different MAR methods on four slices from the clinical test sets with real artifacts, where (a) and (b) are from DeepLesion dataset; (c) and (d) are from Dental dataset. The display window is [-300, 300] HU.}
\label{fig:deepl-dental}
\vspace{-6pt}
\end{figure*}

\subsection{Generalization Comparison on  Clinical Datasets}\label{sec:exp-res-cli}
The ultimate goal is to generalize the MAR networks to clinical CT images corrupted by real metal artifacts. Therefore, we conducted two experiments on the DeepLesion and Dental datasets respectively to evaluate the generalization of different methods on two types of anatomical regions, \ie, torso and dental regions. In either of the two experiments, the training data include paired simulated images and unpaired clinical images in the corresponding dataset, and the test data are those clinical images corrupted by real metal artifacts. 
Note that ground-truths are unavailable for the test data in both experiments. For evaluation, we provide CQA quality result of different MAR methods for quantitative comparison (in the last group of Tables~\ref{tab:torso-quant} and~\ref{tab:dental-quant}), and visualize the predicted MAR results in Fig.~\ref{fig:deepl-dental} for qualitative comparison. In the following, we focus on the clinical MAR performance and omit the results on the simulated DeepLesion dataset for brevity.

Across four typical slices shown in Fig.~\ref{fig:deepl-dental}, the supervised baseline leaves significant metal artifacts that obscure critical anatomical details. Similar to the experimental results in Sec.~\ref{sec:exp-res-syn}, $\beta$-CycleGAN suppresses some metal artifacts. Nonetheless, it introduces novel artifacts and alters the tissue. For example, tissue artifacts can be observed superficial to the clavicle in the image predicted by $\beta$-CycleGAN in Fig.~\ref{fig:deepl-dental}(b). In Fig.~\ref{fig:deepl-dental}(c) and (d), the tissues surrounding external ears are incorrectly removed or deformed by $\beta$-CycleGAN. These behaviors are highly undesirable for clinical practice. 
SemiMAR and UDAMAR demonstrate some improvement, but leaves residual artifacts or spurious cavity when the input images are corrupted by severe metal artifacts. 
In comparison, our \methodname removes both bright and dark streak artifacts induced by metals of different shapes, and restores bony structures (\eg, clavicles in Fig.~\ref{fig:deepl-dental}(b)) and soft tissues (\eg, tongue in Fig.~\ref{fig:deepl-dental}(c) and (d)) that are essential for clinical diagnosis and treatment planning, delivering the most effective artifact reduction. 

Tables~\ref{tab:torso-quant} and~\ref{tab:dental-quant} show that our  \methodname consistently achieves the best CQA quality scores in the clinical datasets (8.55 for clinical DeepLesion and 7.30 for clinical Dental). 
Furthermore, the CQA quality results closely align with the qualitative examples illustrated in Fig.~\ref{fig:deepl-dental}, suggesting that the CQA has the potential to serve as a reliable tool for clinical MAR performance evaluation in the absence of ground-truths.

\subsection{Ablation Studies}\label{sec:exp-abl}
In this subsection, we evaluate the effectiveness of the components in our \methodname, namely the CQA model and the radiologist-in-the-loop self-training. Performance of different configurations of CQA is evaluated on the test set of clinical quality assessment dataset (Secs.~\ref{sec:exp-abl-cqa-train} and \ref{sec:exp-abl-cqa-arch}). Detailed analyses of our \methodname framework (Secs.~\ref{sec:exp-abl-guidance} and \ref{sec:exp-abl-li}) are based on the simulated DeepLesion and CTPelvic1K datasets and follow the same setting as in Sec.~\ref{sec:exp-res-syn}.

\subsubsection{Ablation on the Training of CQA}\label{sec:exp-abl-cqa-train}
\begin{figure}
\centerline{\includegraphics[width=.95\linewidth]{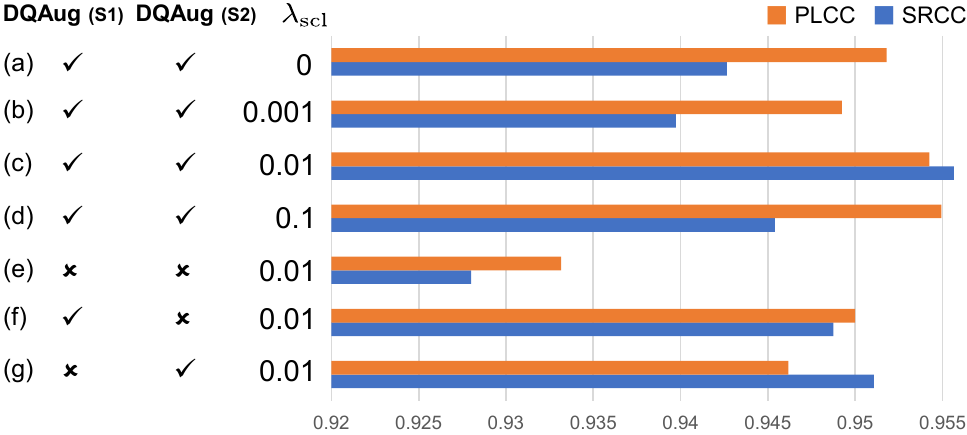}}
\caption{Ablation study on the training of CQA. A checkmark indicates the strategy is applied.}
\label{fig:abl-cqa-train}
\end{figure}

Fig.~\ref{fig:abl-cqa-train} shows the effectiveness of the proposed DQAug strategy and the compound loss function $\mathcal{L}_\mathrm{cqa}(\psi)$ in Eq.~\eqref{eq:cqa-loss}, both of which can impact the performance of CQA through the training process. 

As for the balancing factor $\lambda_\mathrm{scl}$ in $\mathcal{L}_\mathrm{cqa}(\psi)$, the first four rows in Fig.~\ref{fig:abl-cqa-train} show that $\lambda_\mathrm{scl}=0.01$ empirically achieves the best alignment of CQA prediction with radiologists' feedback, compared to other choices. Either diminishing ($\lambda_\mathrm{scl}=0$ or $\lambda_\mathrm{scl}=0.001$) or reinforcing ($\lambda_\mathrm{scl}=0.1$) the contrastive loss will lead to lower SRCC and PLCC values.  
We also explored the effectiveness of two steps in DQAug, \ie, generating moderate-quality images (``DQAug (S1)'') and employing MixUp data augmentation (``DQAug (S2)'') across images in different quality groups. By comparing Rows (c), (e), (f) and (g), we found that DQAug significantly improves the performance of CQA by enhancing the data diversity, enabling CQA to capture subtle quality differences across different artifact-affected CT images.

\begin{figure}
\centerline{\includegraphics[width=\linewidth]{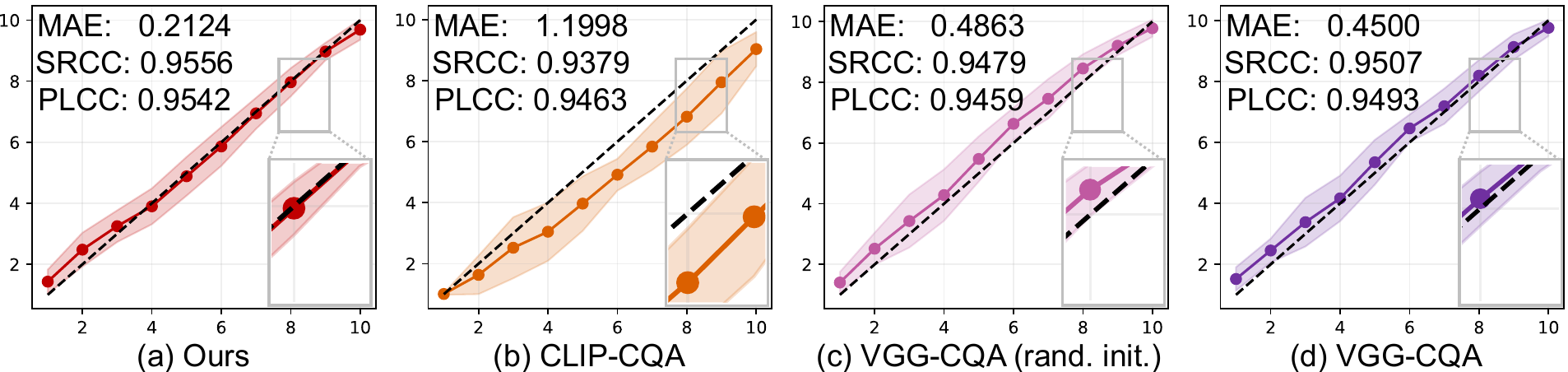}}
\caption{Alignment of predicted quality scores from different CQA architectures with radiologist-annotated scores (black dashed line). Zoom in for better viewing.}
\label{fig:abl-cqa-arch}
\end{figure}

\subsubsection{Ablation on the Design of CQA}\label{sec:exp-abl-cqa-arch}
Fig.~\ref{fig:abl-cqa-arch} investigates alternative models for the radiologist-aligned quality assessment, including VGG19~\cite{vgg} and CLIP-ResNet~\cite{clip}, with our proposed CQA for comparison. 
The motivation for selecting these models is that they are reported to align closely with human visual perception, as they have been pretrained on large-scale image datasets~\cite{perceptual-loss}. 
Each pretrained model is applied as an encoder to extract multiscale features from the input image, and is then integrated with a vectorization layer followed by a quality head, as depicted in Fig.~\ref{fig:method}(b), resulting in two quality assessment models: VGG-CQA and CLIP-CQA. We also compared a variant of VGG-CQA whose parameters are randomly initialized. These models are trained and tested following the same procedure as our CQA. 

Fig.~\ref{fig:abl-cqa-arch} shows that the proposed CQA achieves the best alignment to the radiologists' assessments, with the lowest mean absolute error (MAE) and highest SRCC and PLCC values. Fig.~\ref{fig:abl-cqa-arch}(c) and (d) show that the knowledge from large-scale natural image pretraining offers limited benefits, which might be further enhanced with elaborate adaptation techniques. Despite this, VGG-CQA and CLIP-CQA still underperform, indicating that local features extracted by convolution may not sufficiently capture the radiologists' feedback for overall CT image quality.

\subsubsection{Ablation on the Radiologist-in-the-Loop Self-Training}\label{sec:exp-abl-guidance}
In Table~\ref{tab:abl-mar}, we treat the final developed \methodname 
as the reference model, and then systematically change different components to evaluate their individual contributions to out-of-domain performance (\ie, on the simulated CTPelvic1K dataset), resulting in the following configurations: 
(i) ``w/o $\mathcal{L}_\mathrm{cli}$'': removing the clinical unsupervised loss term in Eq.~\eqref{eq:loss-mar}, which is equivalent to the fully supervised baseline. 
(ii) ``w/o EMA'': keeping the teacher MAR network frozen. 
(iii) ``w/o CQA'': removing the pretrained CQA from the training framework, which is equivalent to setting the quality range in Eq.~\eqref{eq:loss-mar-cli} as $\mathcal{Q}=[1,10]$. 
Furthermore, we explore the optimal $\mathcal{Q}$ for our method by comparing the performance of configurations (iv), (v), (vi), and (vii), which adjust $\mathcal{Q}$ to $[1,4]$, $[4,7]$, $[9,10]$ and $[7,10]$, respectively, with $\mathcal{Q}=[7,10]$ serving as reference.

\begin{table}[t]
\centering
\caption{Ablation study on the training of \methodname. In-domain performance is omitted. The best results are highlighted in \bestres{bold} and the second best results are \secondres{underlined}.}
\resizebox{1.0\linewidth}{!}{
\begin{tabular}{lccc}
\toprule
Configurations & PSNR (dB) & SSIM (\%) & CQA quality \\ 
\midrule
(i) w/o $\mathcal{L}_\mathrm{cli}$ (supervised) & 41.67 & 93.57 &8.154 \\
(ii) w/o EMA &42.15 &94.20 &6.754 \\
(iii) w/o CQA ($\mathcal{Q}=[1,10]$) &45.49 &96.67 &7.575 \\
(iv) $\mathcal{Q}=[1,4]$ &41.99 &93.59 &6.786 \\
(v) $\mathcal{Q}=[4,7]$ &42.39 &93.45 &6.787 \\
(vi) $\mathcal{Q}=[9,10]$ &\secondres{46.55} &\secondres{99.14} &\secondres{8.278} \\
(vii) $\mathcal{Q}=[7,10]$ (reference) &\bestres{47.88} &\bestres{99.25} &\bestres{8.645} \\
\bottomrule
\end{tabular}}
\label{tab:abl-mar}
\end{table}
\begin{figure}[t]
\centerline{\includegraphics[width=\linewidth]{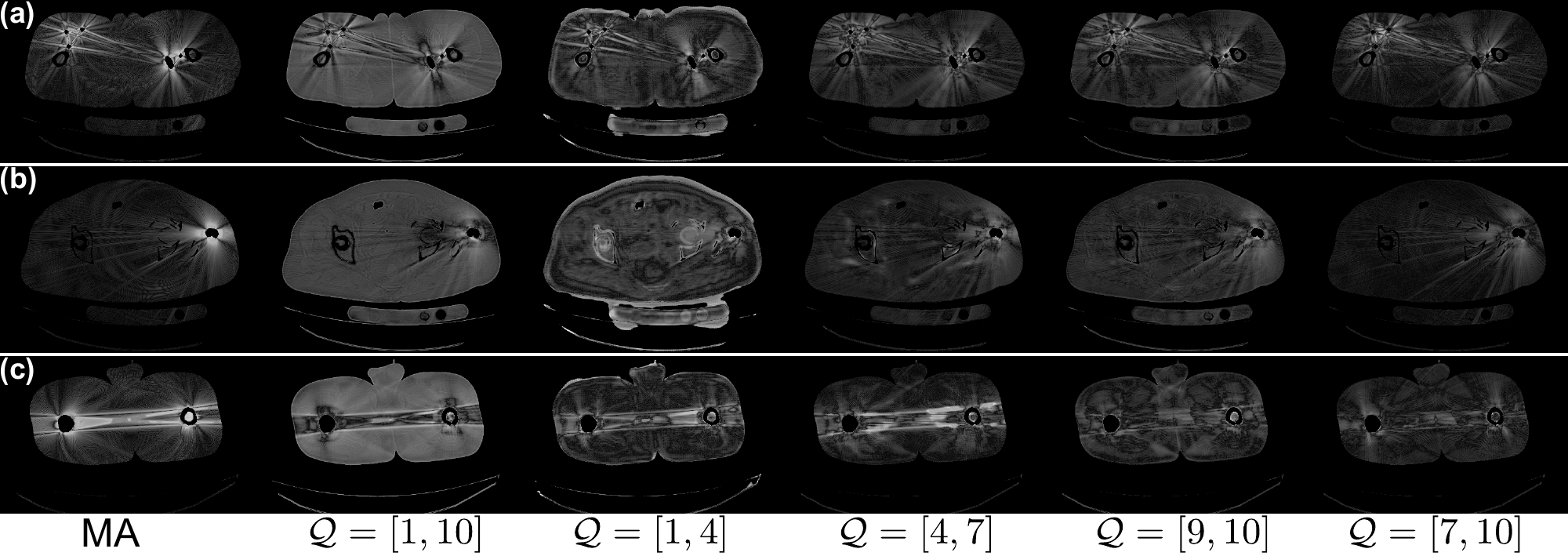}}
\caption{Visualization of absolute error maps of pseudo ground-truths within various quality ranges $\mathcal{Q}$.}
\label{fig:abl-q}
\end{figure}

From Table~\ref{tab:abl-mar}, we observe that removing the unsupervised loss term  $\mathcal{L}_\mathrm{cli}$, EMA update, or clinical quality guidance from CQA, can all significantly degrade the out-of-domain performance. 
In Row (i) of Table~\ref{tab:abl-mar}, the student MAR network lacks supervision from the CTPelvic1K dataset. In Row (ii), the teacher MAR network remains fixed throughout training and relies solely on prior knowledge from simulated MAR pretraining to generate pseudo ground-truths. In this case, the teacher network fails to update its out-of-domain knowledge and provides low-quality pseudo ground-truths for the student network. 
Row (iii) shows improved performance compared to Row (ii), yet the MAR network still suffers from confirmation bias due to training with unfiltered pseudo ground-truths, as evident in the second column of Fig.~\ref{fig:abl-q}. The MAR networks in Rows (iv) and (v) further demonstrate performance degradation when the quality range $\mathcal{Q}$ is constrained to lower values. Interestingly, Row (vi) indicates that using a high quality range (\ie, $\mathcal{Q}=[9,10]$) does not necessarily yield better performance than a slightly broader range ($\mathcal{Q}=[7,10]$).

\begin{figure}[htb]
\centerline{\includegraphics[width=\linewidth]{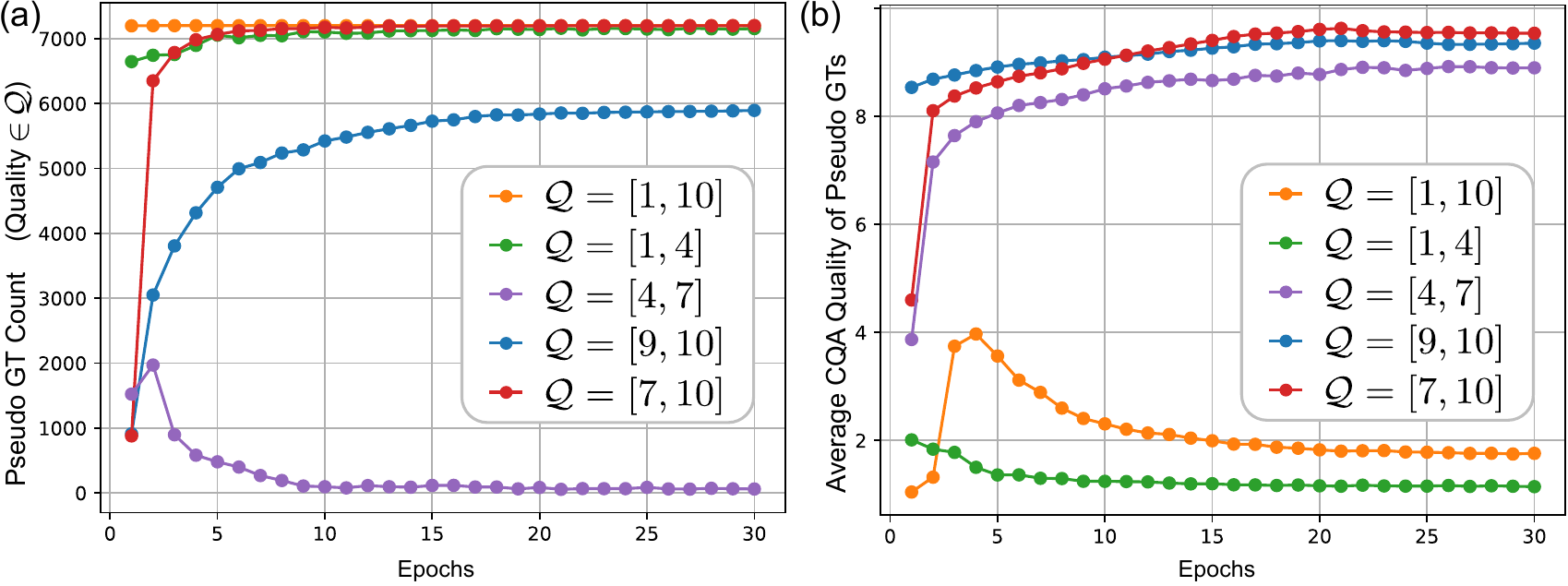}}
\caption{Pseudo ground-truth dynamics during training. (a) Count of pseudo ground-truths within quality range $\mathcal{Q}$. (b) Average CQA quality of all pseudo ground-truths for each $\mathcal{Q}$.}
\label{fig:abl-num-ps}
\end{figure}

To explore this, Fig.~\ref{fig:abl-num-ps} tracks the count of pseudo ground-truths whose CQA quality are within $\mathcal{Q}$ and the average CQA quality of all pseudo ground-truths throughout training.  We found that $\mathcal{Q}=[9,10]$ can be too stringent, excluding many ``moderate-to-high'' quality pseudo ground-truths, leading to a scarcity of out-of-domain data and slightly lower performance than $\mathcal{Q}=[7,10]$. 
With a moderate quality range $\mathcal{Q}=[4,7]$, the pseudo ground-truth count decreases while the average CQA quality increases, suggesting effective self-improvement of generated pseudo ground-truths even with a lower $\mathcal{Q}$, which could be attributed to combined effects of $\mathcal{L}_\mathrm{cli}$ and EMA self-training in our \methodname. 
Additionally, setting a low quality range ($\mathcal{Q}=[1,4]$) or not filtering by quality ($\mathcal{Q}=[1,10]$) leads to low average CQA quality of pseudo ground-truths, negatively impacting performance due to error accumulation from such low-quality pseudo ground-truths. Therefore, we choose $\mathcal{Q}=[7,10]$ for our \methodname.

\subsubsection{Impact of LI-Backprojected CT Images}\label{sec:exp-abl-li}
\begin{figure}[t]
\centerline{\includegraphics[width=\linewidth]{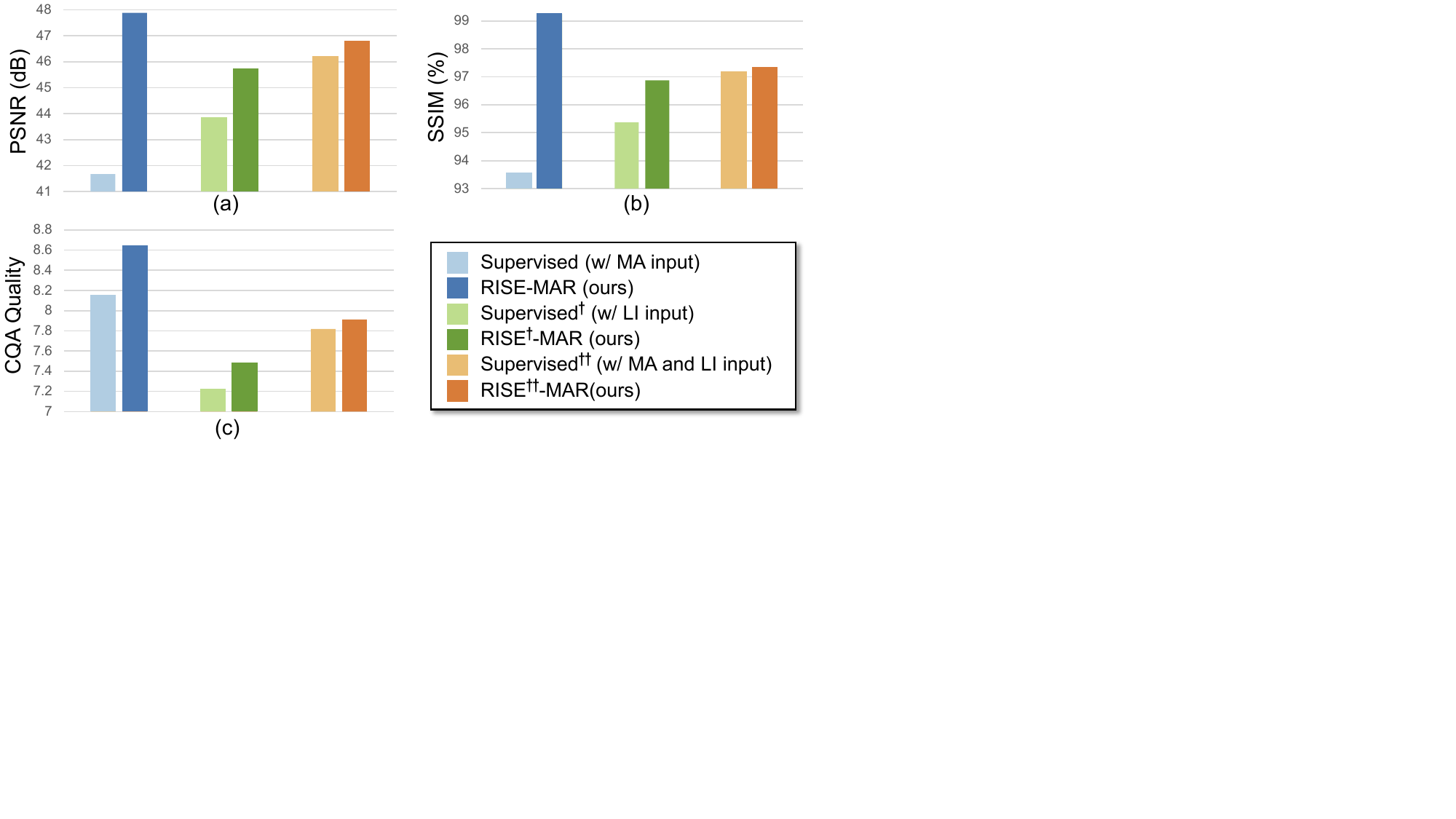}}
\caption{Out-of-domain performance of different variants of \methodname. }
\label{fig:abl-li}
\end{figure}

Fig.~\ref{fig:abl-li} explores whether the proposed framework can be applied to other MAR networks that leverage the information of LI images (\ie, CT images backprojected from linearly-interpolated sinograms). Specifically, we develop two variants from the supervised baseline MAR network, and denote them by ``Supervised$^\dagger$'' and ``Supervised$^{\dagger\dagger}$'', respectively. These MAR networks share similar architectures but differ in their inputs: the vanilla supervised baseline takes the metal artifact-affected CT image as inputs, ``Supervised$^\dagger$'' uses LI images , whereas ``Supervised$^{\dagger\dagger}$'' uses a concatenation of metal artifact-affected images and LI images. Then, we replace the vanilla supervised baseline in our \methodname framework with these variants, resulting in two other frameworks, \methodnameli and \methodnamemali.

The light-colored bars in Fig.~\ref{fig:abl-li} illustrate the performance of supervised baselines, showing notable variations across different input data. 
Incorporating LI images enhances MAR generalization performance on the out-of-domain CTPelvic1K dataset, with the combination of artifact-affected and LI images yielding the best PSNR and SSIM results, as LI images can offer clearer, albeit imperfect, representation of the underlying CT image and benefit the generalization. However, this pattern does not hold for the semi-supervised counterparts. For example, both \methodnameli and \methodnamemali slightly underperform compared to \methodname. In the semi-supervised framework where the MAR network is exposed to \emph{unpaired} out-of-domain data, the presence of secondary artifacts in the LI images may be exaggerated and hinder the model performance. This suggests that effective use of LI images in semi-supervised settings requires more refined approaches. 

Compared to the supervised baselines, the proposed semi-supervised counterparts consistently improve the out-of-domain performance across all metrics. Among them, the MAR network with only metal artifact-affected images as input achieves the largest performance gains, demonstrating superior generalizability enabled by our method.

%% file: secs/4_concl.tex
\section{Discussion and Conclusion}\label{sec:concl}
In this work, we propose \methodname, a simple yet effective semi-supervised framework for clinical MAR application. By building a radiologist-aligned CQA and incorporating it into a self-training framework, we effectively improve the quality and quantity of pseudo ground-truth and produce MAR results that better support clinical diagnosis. 
Extensive experimental results showcase the superiority of \methodname over state-of-the-art methods on real clinical data across different anatomies. We also provide an in-depth analysis on the effectiveness of each component in \methodname, and extend our framework to include other MAR networks that take LI-backprojected~\cite{kalender1987li} CT image as input.

The benefits of incorporating a CQA model that is well aligned with the feedback of radiologists for semi-supervised learning lie in at least two aspects. On the one hand, CQA works as an effective quality filter in pseudo ground-truth generation. This not only alleviates the problem of confirmation bias, but also enhances the disentanglement of metal artifacts and anatomical structures, which is a key point in many unsupervised~\cite{liao2019adn,wang2021bcyclegan} and semi-supervised methods~\cite{wang2023semimar,lyu2021ududonet}, as a high-quality pseudo ground-truth is often less affected by metal artifacts and tissue distortion. On the other hand, CQA also serves as an evaluation tool that provides efficient and radiologist-aligned assessment for a given CT image with different levels of metal artifact corruption. This shows potential in quantitative evaluation of MAR algorithms on real CT images, since artifact-free ground-truths are unavailable. 

We also notice the occasional inconsistencies between the CQA quality score and PSNR/SSIM in evaluating MAR results. This arises from their different focuses. PSNR emphasizes pixel-wise alignment while SSIM evaluates perceived structural similarity, both may overlook subtle anatomical distortions or artifact severity. In contrast, the CQA score prioritizes clinical quality by assessing both the presence of metal artifacts and the preservation of anatomical integrity, as aligned with radiologists' annotations.

Finally, we acknowledge some limitations in our work. First, the alignment of CQA with radiologists' feedback can be further enhanced. In our experiments, the CQA is trained on approximately 50k CT images using the proposed DQAug techniques. While it effectively filters out low-quality images, it may not adequately detect structural distortions when artifacts are not prominent. 
Collecting more CT images with fine-grained quality assessments (\eg, spatial heatmaps highlighting regions with potential errors) annotated by a larger pool of experienced radiologists may address this issue~\cite{iqagpt,liang2024rich}, despite being more time-consuming and labor-intensive. Another limitation is that the quality predicted by CQA is only used to select pseudo ground-truths during training, rather than an explicit objective to guide the updating of the MAR network. Reinforcement learning is a promising technique to address this issue and can be one of our future directions.